\documentclass[aps,showpacs,twocolumn,amsmath,amssymb,superscriptaddress,prl]{revtex4-1}

\usepackage{graphicx}
\usepackage{color}

\usepackage{bm}

\usepackage{comment}

\begin{document}

\title{Investigation of nonlocal transport associated with the orbital Hall effect in Ti}

\author{Keitaro Takashina}
\affiliation{Department of Applied Physics and Physico-Informatics, Keio University, Yokohama 223-8522, Japan}

\author{Naoki Yano}
\affiliation{Department of Applied Physics and Physico-Informatics, Keio University, Yokohama 223-8522, Japan}

\author{Asahi Oe}
\affiliation{Department of Applied Physics and Physico-Informatics, Keio University, Yokohama 223-8522, Japan}

\author{Mari Taniguchi}
\affiliation{Department of Applied Physics and Physico-Informatics, Keio University, Yokohama 223-8522, Japan}

\author{Shuto Kimura}
\affiliation{Department of Applied Physics and Physico-Informatics, Keio University, Yokohama 223-8522, Japan}

\author{Kazuya Ando
\footnote{Correspondence and requests for materials should be addressed to ando@appi.keio.ac.jp}
}
\affiliation{Department of Applied Physics and Physico-Informatics, Keio University, Yokohama 223-8522, Japan}
\affiliation{Keio Institute of Pure and Applied Sciences, Keio University, Yokohama 223-8522, Japan}
\affiliation{Center for Spintronics Research Network, Keio University, Yokohama 223-8522, Japan}

\begin{abstract}
We investigate nonlocal transport in single-layer Ti Hall bars to explore signatures of orbital-current transport driven by the orbital Hall effect. Despite the negligible spin Hall effect in Ti, we observe a finite nonlocal resistance in the single-layer Ti Hall bar and study its dependence on the central channel width. Finite-element simulations show that the measured signal contains a sizable Ohmic bypass contribution. However, the bypass contribution is strongly suppressed at small channel widths and cannot fully account for the observed nonlocal resistance even when variations in the Ti resistivity are taken into account. Our results therefore suggest an additional nonlocal contribution distinct from the Ohmic bypass background, which may be associated with orbital transport driven by the orbital Hall effect in Ti.
\end{abstract}

\maketitle

Understanding and harnessing spin currents have played a central role in the development of spintronics. A key mechanism for generating spin currents is the spin Hall effect (SHE), in which a longitudinal charge current is converted into a transverse spin current through spin-orbit coupling (SOC)~\cite{RevModPhys.87.1213}. The SHE has enabled major advances in spintronics, including the electrical manipulation of magnetization via spin-orbit torque~\cite{RevModPhys.91.035004}, and has provided a versatile route for generating and detecting spin currents in a wide variety of material systems.

While spin currents have been the primary focus of spintronics, electrons in solids possess not only spin but also orbital angular momentum, opening an additional channel for angular momentum transport. Recent studies have established the orbital Hall effect (OHE) as a promising mechanism for generating orbital currents. In contrast to the SHE, the OHE does not inherently require SOC, but instead originates from orbital texture in momentum space and can exist even in light metals~\cite{PhysRevLett.95.066601,PhysRevB.77.165117,PhysRevLett.102.016601,PhysRevLett.121.086602,PhysRevB.98.214405,PhysRevLett.126.056601,PhysRevMaterials.6.095001}. This suggests that the OHE is more ubiquitous than the SHE and may enable efficient angular momentum generation across a broader class of materials.

Experimental signatures of the OHE have been reported using various techniques, such as orbital torque, orbital pumping, and magnetoresistance measurements~\cite{Cr-orbital,Ta-orbital,PhysRevResearch.4.033037,hayashi2022observation,choi2021observation,PhysRevResearch.5.023054,PhysRevLett.125.177201,PhysRevLett.128.067201,PhysRevLett.131.156702,PhysRevB.108.144436,moriya2024nano,BulkSantos2024,gao2024control,hayashi2024observation,hayashi2023orbital,el2023observation,xu2024orbitronics,seifert2023time,taniguchi2025acoustic,ledesma2025nonreciprocity,kashiki2025violation}. These developments have opened new opportunities for angular momentum transport beyond spin-based phenomena and have motivated the rapid development of orbitronics, an emerging field that seeks to understand and harness orbital currents in solid-state systems~\cite{go2021orbitronics,KIM2022169974,jo2024spintronics,2025Andoreview}. 
At the same time, these studies have revealed that the characteristic transport length scale of orbital currents remains an open question~\cite{Liao2026Orbitronics}. Experimental reports have suggested orbital transport over length scales ranging from a few nanometers to several tens of nanometers, whereas theoretical studies have proposed various scenarios, such as ultrashort orbital diffusion lengths caused by orbital quenching and long-range propagation associated with momentum-space hot spots~\cite{PhysRevB.109.214427,PhysRevLett.130.246701,86cm-yn9z}.

In the study of spin transport in materials with long spin diffusion lengths, nonlocal measurements have played an important role. Nonlocal geometries, which spatially separate the generation and detection of angular momentum currents, provide a useful platform for probing lateral transport apart from the primary charge flow. The potentially long transport length of orbital currents suggests that a similar approach is promising for orbital transport. A recent study has indeed employed a nonlocal configuration with a ferromagnetic electrode to investigate orbital transport~\cite{gao2025nonlocal}. Extending this nonlocal approach to single-layer devices may be attractive because the simpler structure enables a more direct examination of orbital transport associated with the OHE while avoiding possible interfacial complications. A similar single-layer strategy has also been explored in spin transport studies. Early work on single-layer Au highlighted the importance of carefully separating background contributions in nonlocal geometries~\cite{mihajlovic2009negative}, and later studies demonstrated clear nonlocal spin Hall signals in Au and Cu films $\delta$-doped with Bi~\cite{PhysRevLett.122.016804}.

\begin{figure}[tb]
\includegraphics[scale=1]{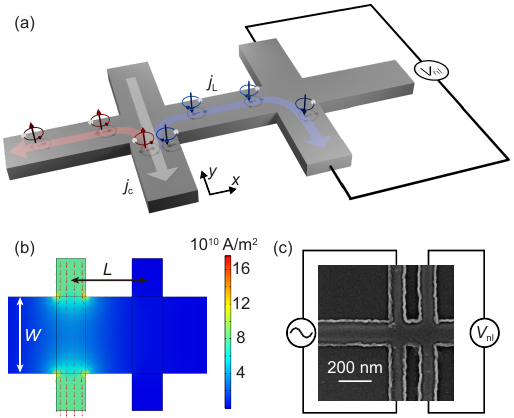}
\caption{
(a) Schematic of the nonlocal measurement geometry for orbital transport. A charge current $j_{\mathrm{c}}$ is injected through the left terminal along the $y$ axis, which generates an orbital current $j_{\mathrm{L}}$ flowing along the $x$ axis in the central channel via the OHE. The orbital current is converted into a charge current along the $y$ axis through the inverse OHE, giving rise to a nonlocal voltage $V_{\mathrm{nl}}$ detected at the right terminals. 
(b) Representative current-density distribution in the Ti Hall bar calculated using COMSOL Multiphysics, explicitly taking into account the finite terminal widths. The color scale and arrows represent the magnitude and local flow direction of the current density, respectively. A fraction of the injected charge current flows directly into the detection branch because of the device geometry, producing a nonlocal voltage unrelated to orbital transport, which is the Ohmic bypass contribution. The central channel width and length are denoted by $W$ and $L$, respectively.
(c) A scanning electron microscopy (SEM) image of the Ti Hall bar device used for the nonlocal measurement. 
}
\label{nonlocal} 
\end{figure}

In this Letter, we investigate nonlocal transport associated with the orbital Hall effect in single-layer Ti Hall bars. Ti provides an attractive platform because it is expected to host a sizable OHE, while spin-related contributions are negligible owing to its extremely small SHE~\cite{PhysRevMaterials.6.095001,hayashi2022observation,choi2021observation}. Figure~\ref{nonlocal}(a) shows a schematic of the device structure and measurement geometry. When a charge current is applied to the current injection terminal along the $y$ axis, the OHE generates orbital currents that flow along the $x$ axis. These orbital currents are converted into charge currents along the $y$ axis through the inverse OHE, giving rise to a nonlocal voltage at the detection electrodes. A central challenge in this geometry is the presence of an Ohmic bypass contribution arising from the direct current path between the injection and detection electrodes (see Fig.~\ref{nonlocal}(b)). We show that analyzing the dependence of the nonlocal signal on the central channel width $W$ provides an effective route to evaluate this bypass contribution. The observed nonlocal signal in single-layer Ti cannot be fully explained by the bypass contribution alone, suggesting a contribution associated with the OHE.

Figure~\ref{nonlocal}(c) shows a scanning electron microscopy (SEM) image of the Ti Hall bar. The Ti nonlocal devices were patterned by electron-beam lithography and fabricated by lift-off after the deposition of a 30-nm-thick Ti layer and a 4-nm-thick SiO$_2$ capping layer. All samples were deposited on thermally oxidized Si substrates at room temperature using RF magnetron sputtering. The base pressure of the sputtering chamber was below $6 \times 10^{-7}\,\text{Pa}$ to minimize oxidation and impurities during the deposition of both the Ti layer and the SiO$_2$ cap. The resulting Ti Hall bars, with central channel widths $W$ ranging from 80 to 180 nm, were integrated with photolithographically defined Ti/Au contact electrodes, with thicknesses of 3nm for Ti and 150nm for Au. The center-to-center distance between the injection and detection terminals is $L=200$~nm.

Nonlocal transport measurements were performed by applying an AC current with amplitude $I$ and frequency $f=131\,\text{Hz}$ to the current injection terminal using a Keithley 6221 current source. The first-harmonic nonlocal voltage $V_\mathrm{nl}$ between the electrodes of the voltage detection terminal was measured using a Stanford SR830 DSP lock-in amplifier. The first-harmonic nonlocal resistance $R_{\mathrm{nl}}$ was determined from the slope of the linear dependence of $V_{\mathrm{nl}}$ on $I$.

\begin{figure}[tb]
\includegraphics[scale=1]{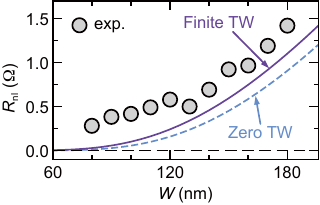}
\caption{
Measured nonlocal resistance $R_{\mathrm{nl}}$ as a function of the central channel width $W$. The gray circles represent the experimental data. 
The dashed curve represents the calculated bypass contribution using Eq.~(\ref{bypass}), which assumes negligible terminal widths (TWs) with respect to $L$.
The solid curve shows the simulated bypass contribution accounting for finite TWs using a Ti resistivity of $70\,\mu\Omega\mathrm{cm}$ determined from sheet-resistance measurements of a reference film (a Ti film fabricated under the same sputtering conditions on a different substrate).
}
\label{COMSOL-exp} 
\end{figure}

Figure~\ref{COMSOL-exp} shows the measured nonlocal resistance $R_\mathrm{nl}$ as a function of the channel width $W$. $R_\mathrm{nl}$ remains finite over the entire measured range of $W$ from 80 to 180 nm. An important contribution to the measured nonlocal resistance $R_\mathrm{nl}$ is the Ohmic bypass signal $R^{\text{by}}_\text{nl}$. Assuming that the terminal widths are negligible compared with the channel length $L$, the bypass contribution is given by
\begin{equation}
  R^{\text{by}}_\text{nl} = \frac{4}{\pi} R_{\text{sq}} \exp \left( -\frac{\pi L}{W} \right),
  \label{bypass}
\end{equation}
which is obtained from the solution of the Laplace equation for a metallic Hall bar~\cite{abanin2011giant}. As shown in Fig.~\ref{COMSOL-exp}, the measured $R_\mathrm{nl}$ is clearly larger than the calculated bypass contribution using Eq.~(\ref{bypass}) (dashed curve).

The difference between the measured $R_\mathrm{nl}$ and the calculated bypass contribution suggests an additional contribution superimposed on the Ohmic background, which may include SHE- and OHE-related signals. Using the reported spin Hall angle of Ti, $\theta_{\mathrm{SH}} = -3.6 \times 10^{-4}$~\cite{du2014systematic}, we estimate that the nonlocal resistance due to the SHE is orders of magnitude smaller than that observed experimentally, suggesting a possible role of the OHE in $R_\mathrm{nl}$. To examine this interpretation more reliably, however, it is necessary to evaluate the bypass contribution exceeding the limitations of the simplified model of Eq.~(\ref{bypass}), which assumes negligible terminal widths. We therefore performed finite-element simulations using COMSOL Multiphysics, explicitly incorporating the finite terminal widths to evaluate the current distribution in the Hall bar geometry. An example of the simulated current density distribution is shown in Fig.~\ref{nonlocal}(b).

In Fig.~\ref{COMSOL-exp}, we compare the simulated bypass contribution $R^{\text{by}}_\text{nl}$ (solid curve) with the approximate result obtained from Eq.~(\ref{bypass}) (dashed curve). The simulation yields a slightly larger $R^{\text{by}}_\text{nl}$ than the approximate expression over the entire range of $W$, highlighting the importance of taking the finite terminal widths into account in the analysis of nonlocal measurements. Nevertheless, the simulated $R^{\text{by}}_\text{nl}$ remains smaller than the experimentally observed $R_\mathrm{nl}$. In particular, the finite nonlocal resistance observed at smaller $W$ cannot be explained by the simulated bypass signal, which is strongly suppressed in this regime. This discrepancy from the Ohmic simulation indicates that the measured resistance cannot be attributed solely to the bypass effect. It is consistent with a scenario involving a finite OHE-induced contribution, where the injected charge current is converted into a transverse orbital current via the OHE, which subsequently produces a measurable nonlocal signal across the voltage detection terminals.

\begin{figure}[tb]
\includegraphics[scale=1]{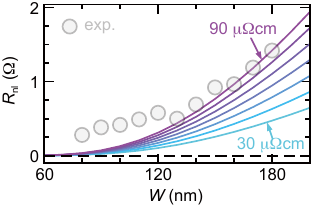}
\caption{
Colored curves show simulated simulated bypass contribution as a function of the central channel width $W$ for Ti resistivities ranging from $30\,\mu\Omega\mathrm{cm}$ to $90\,\mu\Omega\mathrm{cm}$ in steps of $10\,\mu\Omega\mathrm{cm}$. Experimental data are also plotted for comparison. The simulation shows that the bypass signal is strongly suppressed as $W$ decreases, whereas the experimental values remain systematically larger than the simulated bypass contribution in the small-$W$ regime.
}
\label{bypass_sim}
\end{figure}

To assess the robustness of the bypass signal calculation, we performed additional simulations allowing for possible deviations of the Ti resistivity from the measured value of $70\, \mu\Omega \text{cm}$, as shown in Fig.~\ref{bypass_sim}. Despite this parameter sweep, none of the simulated bypass signals reproduced the observed nonlocal resistance in the small-$W$ regime, where the bypass contribution is exponentially suppressed. This discrepancy between the experimental data and the Ohmic model indicates that the nonlocal signal cannot be attributed solely to electronic leakage. Given that the SHE contribution is negligible in this system, the remaining signal could potentially suggest long-range orbital transport in Ti.

In conclusion, we investigated the nonlocal resistance of the Ti Hall bar devices without any magnetic layer. We observed a systematic dependence of the nonlocal resistance on the central channel width $W$ of the Hall bar. Finite-element analysis revealed a sizable bypass contribution arising from the device geometry, particularly from the finite terminal widths. However, even after taking this geometric effect into account, the bypass contribution is strongly suppressed as $W$ decreases and cannot explain the observed nonlocal resistance, especially in the small-$W$ regime. This discrepancy suggests an additional nonlocal contribution distinct from the Ohmic background, which may be associated with orbital transport driven by the OHE. A more conclusive understanding of nonlocal orbital transport will require a quantitative model for this geometry, which is beyond the scope of the present study. The present results provide a basis for further experimental and theoretical investigation.

\begin{acknowledgments}
    This work was supported by JSPS KAKENHI (Grant Numbers: 22H04964, 25K21707), Spintronics Research Network of Japan (Spin-RNJ), and MEXT Initiative to Establish Next-generation Novel Integrated Circuits Centers (X-NICS) (Grant Number: JPJ011438). 
\end{acknowledgments}


\end{document}